\documentclass[aps,prb,twocolumn,groupedaddress]{revtex4}

\usepackage{graphicx}
\usepackage{dcolumn}
\usepackage{epsfig}
\usepackage{bm}
\usepackage{amsfonts}
\usepackage{latexsym}

\newcommand{\be}{\begin{equation}}
\newcommand{\ee}{\end{equation}}
\newcommand{\bea}{\begin{eqnarray}}
\newcommand{\eea}{\end{eqnarray}}

\newcommand{\figwidth}{0.75\columnwidth}

\begin{document}


\title{Two-Hole and Four-Hole Bound States in a $t-J$ Ladder at half-filling}


\author{Weihong Zheng}
\email[]{w.zheng@unsw.edu.au}
\homepage[]{http://www.phys.unsw.edu.au/~zwh}
\affiliation{School of Physics,
The University of New South Wales,
Sydney, NSW 2052, Australia.}

\author{C.J. Hamer}
\email[]{c.hamer@unsw.edu.au}
\affiliation{School of Physics,
The University of New South Wales,
Sydney, NSW 2052, Australia.}


%

\date{\today}

\begin{abstract}
The two-hole excitation spectrum of the $t-J$ ladder at half-filling is studied
using linked-cluster series expansion methods. A rich spectrum of bound states emerges,
particularly at small $t/J$. Their dispersion relations and coherence lengths are
computed, along with the threshold behaviour as the bound states merge into the
continuum. A class of 4-hole bound states is also studied, leading to the conclusion that
phase separation occurs for $t/J \lesssim 0.5$, in agreement with other studies.
\end{abstract}


\maketitle

\section{\label{sec:intro}INTRODUCTION}

Ladder models of magnetic and electronic systems, such as
the Heisenberg, $t-J$ and Hubbard ladders, have attracted great interest
in recent years \cite{dag96,ric97}, on both theoretical and experimental grounds. 
The ladders provide a ``half-way house" between one-dimensional and 
two-dimensional systems, and display new and interesting behaviour in their
own right. The spin-$\frac{1}{2}$ Heisenberg ladder, for instance,
forms a magnetically disordered {\it spin liquid}, with a finite spin gap\cite{dag96}.
Experimentally, two-leg $S=\frac{1}{2}$ ladders are found in some cuprates like
SrCu$_2$O$_3$\cite{azu94}; while the ladder material Sr$_{14-x}$Ca$_x$Cu$_{24}$O$_4$
has recently been found to show superconductivity under high pressure\cite{ueh96}.

In this paper we study two-hole bound states which occur in the $t-J$ ladder
at half-filling, using a newly developed linked-cluster series expansion method
for 2-particle states\cite{tre00}. A rich spectrum of multiparticle bound states
will be revealed. The properties of 2-particle states, such as their dispersion relations,
coherence lengths and exclusive structure factors, can provide much important information about
the dynamics of the model at hand, providing a stringent  test of any analytical
model of the system. These properties are also amenable to experimental test, via
neutron scattering or other techniques. In the case of the Heisenberg ladder, for instance,
a model\cite{kot98,sus98} based on triplet excitations of a ground state 
consisting of singlet dimers on each rung of the ladder\cite{sac90,gop94,Dam98}
provides  a reasonable qualitative description of the numerical results for both 1- and 2-particle
excitations, not only in the dimer limit, but even for the isotropic case.
Ideally, one would like to achieve a similar convergence between
analytical theory and numerical  `experiment' for the case of the $t-J$ ladder.

The basic properties of the $t-J$ ladder have been explored by exact 
diagonalizations\cite{dag92,poi95,tro96,hay96,hay96b,mul98,rie99},
the density renormalization group\cite{whi97,sie98,gaz99,rom00},
quantum Monte Carlo simulation\cite{bru01}, and series expansion techniques\cite{oit99,zwh02},
as well as approximate analytic theories\cite{sie98,gaz99,sig94,lee,sus99,jur01}.
The Hamiltonian of the $t-J$ ladder is
\begin{widetext}
\begin{eqnarray}
H = && J \sum_{i,a} ( {\bf S}_{i,a} \cdot {\bf S}_{i+1,a}
   -{1\over 4} n_{i,a} n_{i+1,a} ) 
  + J_{\perp} \sum_{i} ( {\bf S}_{i,1} \cdot {\bf S}_{i,2}
   -{1\over 4} n_{i,1} n_{i,2} ) \nonumber \\
 && - t \sum_{i,a,\sigma} P (c^{\dag}_{i,a,\sigma} c_{i+1,a,\sigma}
        + H.c. ) P  
  - t_{\perp} \sum_{i,\sigma} P (c^{\dag}_{i,1,\sigma} c_{i,2,\sigma}
        + H.c. ) P  \label{eqH}
\end{eqnarray}
\end{widetext}
where $i$ labels sites along each chain, $\sigma$ ($=\uparrow$ or
$\downarrow$) and $a$ (=1,2) are spin and leg indices,
 and $P$ is a projection
operator which excludes doubly occupied sites. The constants
$J$, $t$ are exchange and hopping parameters on each chain,
while $J_{\perp}$, $t_{\perp}$ are coupling parameters
between the two chains, i.e. on the rungs of the ladder.

The scenario is then as follows\cite{tro96}.

Starting from the half-filled case, and in the dimer limit of strong interchain coupling
$J_{\perp}, t_{\perp}$, the ground state consists of spin singlet dimers on each rung,
as in the Heisenberg ladder. The lowest spin excitation is a triplet excitation on one rung,
propagating via the coupling between the rungs. There are also hole excitations with 
spin-$\frac{1}{2}$, which carry both spin and charge. The lowest 2-hole excitation consists
of a singlet hole pair on one rung, which again develops into a band by propagation along the
ladder. Relative to the 2-hole ground state, this then corresponds to a gapless band of particle-hole charge
excitations: thus the system is said to be in a Luther-Emery C1S0 phase (1 gapless charge mode,
0 gapless spin modes). The spin gap evolves discontinuously away  from half-filling\cite{tro96},
because there is a triplet particle-hole excitation of lower energy than the triplet
magnon state referred to above.

A phase diagram in the $J/t$ versus electron density $n$ plane, for the case of isotropic
couplings, has been proposed by Poilblanc {\it et al.}\cite{poi95} and by M\"uller and
Rice\cite{mul98}. For $J/t \gtrsim 2$, phase separation is predicted to occur\cite{rom00}.
For $J/t\lesssim 2$ and low to moderate doping the system is in a C1S0 phase, and crosses 
to a C1S1 phase (i.e. the spin gap vanishes) for higher doping. For small $J/t$ and low doping
M\"uller and Rice\cite{mul98} also find ferromagnetic (Nagaoka) and C2S2 phases. Superconducting
pairing correlations occur away from half-filling\cite{dag92,hay96,sig94}.

Jurecka and Brenig\cite{jur01} have recently constructed an analytic model of the 2-hole bound states on the $t-J$
ladder, starting from a rung dimer ground state as discussed above, and including
both 1-hole pseudofermion elementary excitations and 2-hole singlet excitations on a single rung. They find
a low-lying $S=0$ bound state in good agreement with exact diagonalization calculations, but the
triplet gap is underestimated as the isotropic limit is approached.

The outline of the paper as follows: in Sect. II we briefly  describe 
our series expansion method, while
Section III  present results for pairs of  `even parity' and `odd parity'
excitations respectively, A rich and complex spectrum of bound states is revealed.
Sect. IV discusses 4-hole bound states formed from two hole-pairs, each on a single rung,
and this allows some discussion of phase separation.
Sect. V give a summary and discussion. 
All the bound states discussed in this paper involve identical `single
particle' components; bound states of non-identical components will be
treated in a separate paper.


\section{\label{sec:method}Method}

We employ a
``rung basis", in which the 2nd and 4th terms in (\ref{eqH})
form the unperturbed Hamiltonian $H_0$, and the remaining terms are
treated perturbatively. That is, the Hamiltonian (\ref{eqH}) 
is rewritten as
\be
H = H_0 + x V
\ee
where 
\bea
H_0 &=&  J_{\perp} [\sum_{i} ( {\bf S}_{i,1} \cdot {\bf S}_{i,2}
   -{1\over 4} n_{i,1} n_{i,2} ) \nonumber \\
 &&  - y_1 \sum_{i,\sigma} P (c^{\dag}_{i,1,\sigma} c_{i,2,\sigma} 
        + H.c. ) P ]
\eea
\bea
V &=& J_{\perp} [ \sum_{i,a} ( {\bf S}_{i,a} \cdot {\bf S}_{i+1,a}
      -{1\over 4} n_{i,a} n_{i+1,a} )  \nonumber \\
   && - y_2 \sum_{i,a,\sigma} P (c^{\dag}_{i,a,\sigma} c_{i+1,a,\sigma} 
        + H.c. ) P  ]
\eea
and $x=J/J_{\perp}$, $y_1=t_{\perp}/J_{\perp}$, and $y_2 = t/J$.

The eigenstates of $H_0$ are direct products constructed
from the nine possible rung states, which are listed in Table I for
easy reference.
In the half-filled case, the lowest energy rung state is
 a spin-singlet state 
 $ (\mid \uparrow \downarrow \rangle - \mid \downarrow \uparrow \rangle )/\sqrt{2}$.
The ground state of the unperturbed Hamiltonian,
at half-filling, is then a direct product of these spin-singlet states on each
rung. The lowest spin excitation, at half-filling, consists of a spin-triplet
excitation on one rung, which propagates coherently along the ladder. 
One and
two-hole charge excitations are created by removing one or two electrons
from a rung state and allowing these to propagate, via the $t$-hopping
term. 
In our previous work\cite{oit99}, we have studied the 1-hole properties, and 
the properties of 2-holes sitting on the same rung.
The study of the dynamic properties of 2-holes sitting on different rungs
is much more complicated: these have been studied by Jurecka and Brenig\cite{jur01}
and by Troyer,  Tsunetsugu and  Rice\cite{tro96}.

Recently, we have developed strong-coupling series expansion methods to 
study two-particle spectra of quantum lattice models\cite{zwh01}.
These methods allow us to 
precisely determine the low-lying excitation spectra of the models at hand,
including all two-particle bound/antibound states, at least for small $x$.
In this paper, we apply this new technique to study the dynamic
properties of 2-holes in the $t-J$ ladder.

As discussed in our previous paper\cite{zwh01}, to compute the two-particle properties,
we first calculate an effective Hamiltonian in the two-particle sector
\begin{equation}
E_{2}({\bf i,j;k,l})=\langle {\bf k,l} | H^{\rm eff} | {\bf i,j} \rangle \;,
\end{equation}
and then calculate the irreducible two-particle matrix element
\begin{eqnarray}
\Delta_{2}({\bf i,j;k,l}) &  = & E_{2}({\bf i,j;k,l}) -E_{0}(\delta_{{\bf
i,k}}\delta_{{\bf j,l}} + \delta_{{\bf i,l}}\delta_{{\bf j,k}}) \nonumber \\
&& -
\Delta_{1}({\bf i,k})\delta_{{\bf j,l}} - \Delta_{1}({\bf i,l})
\delta_{{\bf j,k}} \nonumber\\
 && - \Delta_{1}({\bf j,k}) \delta_{{\bf i,l}} -
\Delta_{1}({\bf j,l}) \delta_{{\bf i,k}} \;,
\end{eqnarray}
where $\delta$ refers to a Kronecker delta function and
$\Delta_{1}$ is the one-particle irreducible matrix element.
Once the effective two-particle Hamiltonian is known, we still have to solve the 
Schr\"odinger equation.
The two particle continuum is delimited by the maximum (minimum)  energy of
two single particle excitations whose combined momentum is the center
of mass momentum. 
There may be multiple solutions to the Schr\"odinger equation
above or below the two-particle continuum. 
Those solutions with energy
below the bottom edge of the continuum are the bound states, while those with energy
higher than the upper edge of the continuum are the antibound states. The binding energy
is defined as the energy difference
between the lower edge of the continuum and the energy of the bound state,
while the antibinding energy is defined as the energy difference
between the upper edge of  continuum and the energy of an antibound state.

To compute the perturbation series we choose values for the free
parameters $y_1, y_2$, and derive expansions in powers of $x$. 
Series have been computed to order $x^{11}$. 
The calculations involve (trivial) 1-dimensional
clusters up to 12 rungs, and are limited to this order by computer
memory constraints.  
To avoid excessive data here we mainly consider the case that $y_1=y_2$.
The series for the energies of the lowest singlet bound states $S_1$ and $S_2$ 
and triplet bound state $T_1$ (and also the lower edge  of the 
continuum) at band maximum  $k=\pi$ for $y_1=y_2=0.1$
are given in Table \ref{tab_ser}. Other series
are available on request.

In the small $x$ limit our strong coupling series are
highly accurate and we can find all details of various two-particle
bound (and antibound) states. The overall 2-particle spectrum is
much richer than that obtained in previous studies.
Several singlet and triplet bound/antibound states  are found. 
The number of bound states depends on the coupling constants
as well as the wavevector.

Previous numerical studies\cite{tro96,jur01} on small lattices
have concentrated on the coupling
$y_1=1/3$, $y_2=10/3$, and $t=t_{\perp}$, and found one singlet
bound state in the even parity channel. Unfortunately, our series
expansions do not converge well at this coupling, so we can not
make a direct comparison. In order to check the correctness of our results,
we also did some finite lattice calculations on small systems, and find
very good agreement for the lowest energy bound state. The 
finite lattice calculations on small systems certainly are not accurate enough,
however,  to see
other bound states sitting very close to the lower edge of the continuum.

\section{\label{sec:2holes}two-hole bound states}

\subsection{\label{sec:evenparity}Even-parity channel}

\begin{figure}[htb]
  { \centering
    \includegraphics[width=\figwidth]{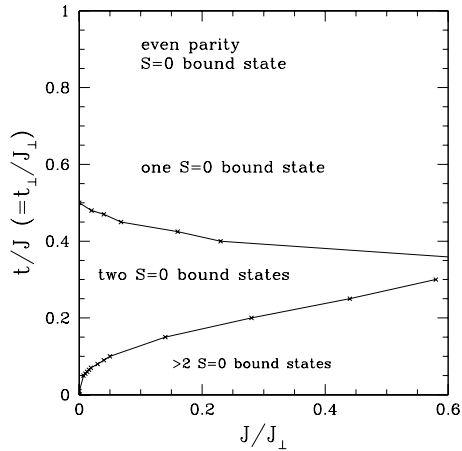}
    \caption{
A diagram for the number of singlet bound states for the even-parity channel in the plane of 
$t/J(=t_{\perp}/J_{\perp})$ and $J/J_{\perp}$.
    \label{fig_phadia_even_S0}} 
  }
\end{figure} 

\begin{figure}[hbt]
  { \centering
    \includegraphics[width=\figwidth]{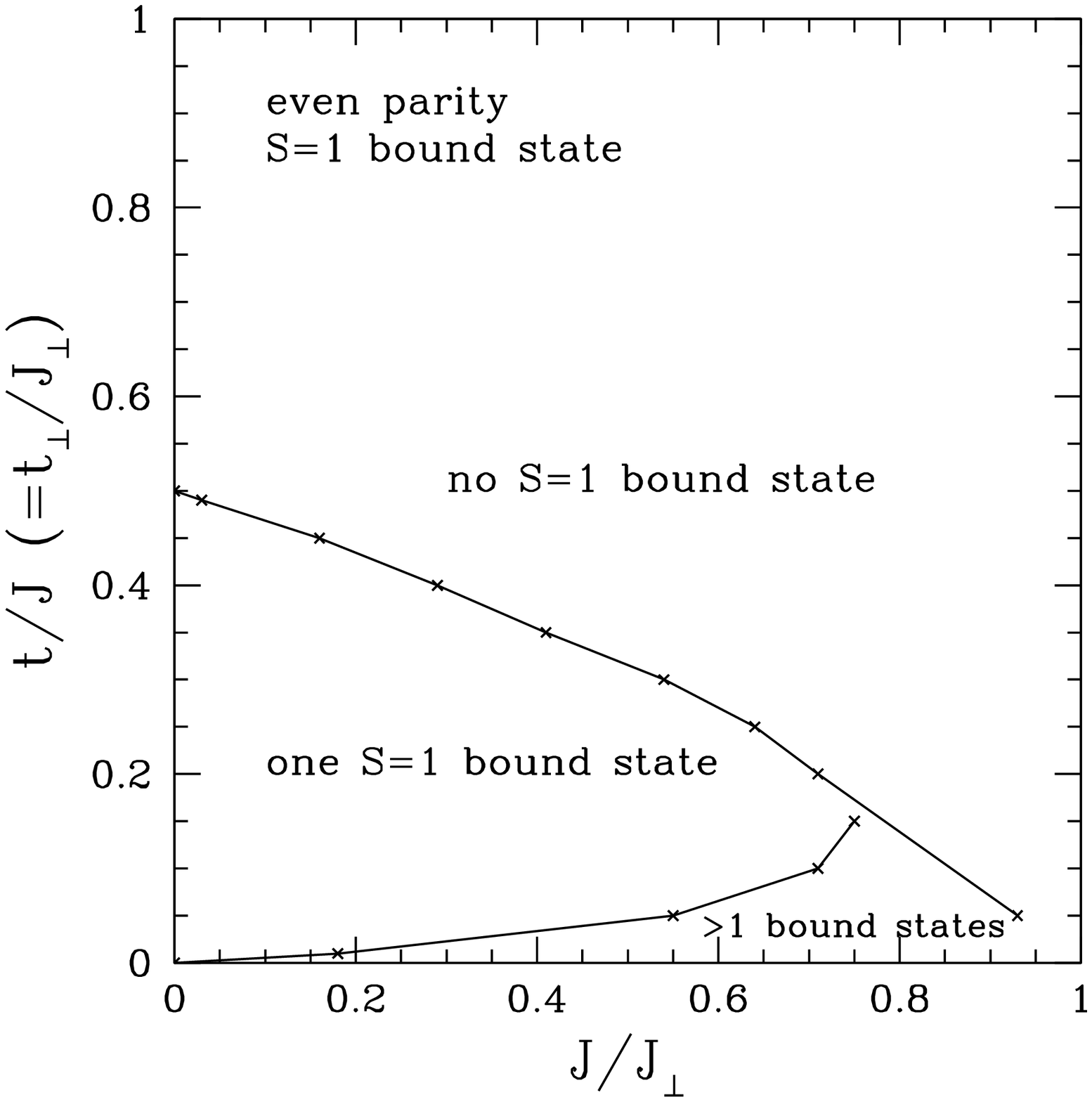}
    \caption{
    A diagram for the number of triplet bound states for 
    the even-parity channel in the plane of 
$t/J(=t_{\perp}/J_{\perp})$ and $J/J_{\perp}$.
    \label{fig_phadia_even_S1}} 
  }
\end{figure} 






Among the nine eigenstates 
for the unperturbed Hamiltonian $H_0$ on each rung, there are the following
two spin-${1\over 2}$  electron-hole bonding states\cite{zwh01}
\bea
&&\frac{1}{\sqrt 2}(\mid 0 \downarrow \rangle + \mid \downarrow 0 \rangle )\nonumber \\
&& \\
&&\frac{1}{\sqrt 2}(\mid 0 \uparrow \rangle + \mid \uparrow 0 \rangle ) ~, \nonumber
\eea
which have `even parity' with respect to reflection across the ladder.
The eigenenergy for these two states is $-t_{\perp}$.
In this section, we discuss the bound states formed from two of these excitations. 

We find that the number of bound states and antibound states depends
strongly on $J/J_{\perp}$ and $t/J (=t_{\perp}/J_{\perp}$). Ignoring the antibound states,
Fig. \ref{fig_phadia_even_S0} show the number of singlet ($S=0$)
bound states in the plane of $t/J(=t_{\perp}/J_{\perp})$ and $J/J_{\perp}$.
We can see that for large $t/J$ ($t/J \gtrsim 0.5$)
there is only one singlet bound state,  while around $t/J\simeq 0.3$,
there is a region where there are two singlet bound states, and 
for smaller $t/J$ and
$J/J_{\perp}>0$, there is a region where there are more than 2 singlet bound states.
In this region, the number of bound state increases as  $J/J_{\perp}$ increases.
Note that all these are in addition to the singlet state consisting of
a pair of holes on the same rung.

Fig. \ref{fig_phadia_even_S1} shows the  number of triplet ($S=1$)
bound states in the same plane.
We can see that for large $t/J$ ($t/J \gtrsim 0.5$)
there are no triplet bound states, whereas around $t/J\simeq 0.2$ and small $J/J_{\perp}$,
there is a region where there is only one triplet bound state, and for smaller $t/J$ and
$J/J_{\perp}>0$, there is a region where there are multiple triplet bound states.
The existence of these
multiple bound states has not been noted before, to our knowledge. Note also
that the triplet bound states appear to vanish as 
the isotropic case $J/J_{\perp}$ is approached.

\begin{figure}[tbh]
  { \centering
    \includegraphics[width=\figwidth]{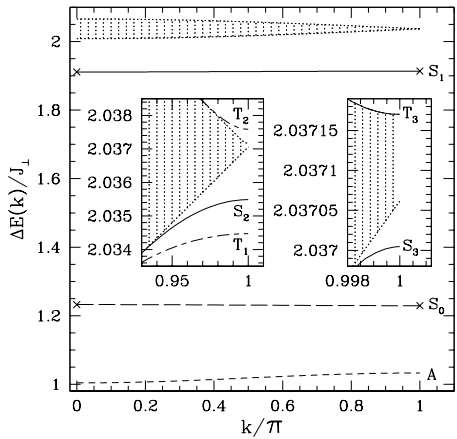}
    \caption{
    The excitation spectrum for the even parity channel with
          $t/J=t_{\perp}/J_{\perp}=0.1$ and $J/J_{\perp}=0.2$. 
          Beside the two-particle
          continuum (gray shaded), there are
          three singlet bound states ($S_1$, $S_2$ and $S_3$) and
          one triplet bound state ($T_1$) below the continuum, and two
          triplet antibound states ($T_2$ and $T_3$) above the continuum.
          The insets enlarge the region near 
          $k=\pi$ so we can see $S_2$, $S_3$, $T_1$ and $T_2$,
           but the antibound state
          $T_3$ is still indistinguishable from the continuum.
           Curve  A is the dispersion of the one-hole bonding state, 
           while curve  $S_0$ is the dispersion of
          two-holes sitting on the same rung. The cross points
          are the results of exact diagonalization for the 8-rung finite lattice 
          with periodic
          boundary condition.
     \label{fig_mk_even_yp1_xp2}}  }
\end{figure} 

\begin{figure}[bth]
  { \centering
    \includegraphics[width=\figwidth]{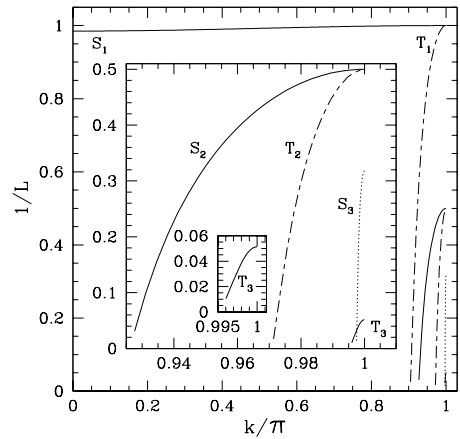}
    \caption{The inverse of the coherence length $1/L$ 
          versus momentum $k$ for three singlets ($S_1$, $S_2$ and $S_3$), 
          one triplet ($T_1$) bound state, and two triplet ($T_2$ and $T_3$) 
          antibound states  for the even parity channel
          at $t/J=t_{\perp}/J_{\perp}=0.1$ and $J/J_{\perp}=0.2$. 
          The inset enlarges the region near 
          $k=\pi$.
     \label{fig_CL_even_yp1_xp2}}  }
\end{figure}  

Now let us look at a particular case 
 $t/J=t_{\perp}/J_{\perp}=0.1$ (i.e. small hopping parameters) 
 and $J/J_{\perp}=0.2$.
The two-particle excitation spectrum is shown 
in Fig. \ref{fig_mk_even_yp1_xp2}.
One can see from these graphs 
that there are three singlet bound states ($S_1$, $S_2$ and $S_3$) and
one triplet bound state ($T_1$) below the continuum, and two
triplet antibound states ($T_2$ and $T_3$) above the continuum,
in addition to the state $S_0$ corresponding to a hole pair on a single
rung. All the dispersion bands are very flat, as expected for small
hopping parameters.
The singlet bound state $S_1$ exists for the whole range 
of momenta (its coherence length $L$ is finite also for the whole range 
of momenta), while other
bound/antibound states exist only in a limited range of momenta near $k=\pi$. 
Figure \ref{fig_CL_even_yp1_xp2} shows the inverse of the coherence length $1/L$ 
versus momentum $k$ for these various bound states.
At $k=\pi$, the coherence length $L$ for $S_1$ and $T_1$ is 1,  for
$S_2$ and $T_2$, it is about 2, $S_3$ has coherence length $L$ about 3, while
for $T_3$ the coherence length $L$ is very large, about 20 (its antibinding energy at $k=\pi$
is very small, $E_b/J_{\perp}=8.9\times 10^{-8}$, so $T_3$ is indistinguishable from 
the two-particle continuum in Fig. \ref{fig_mk_even_yp1_xp2}).
This means that the formation of bound/antibound
 states $S_1$, $T_1$ , $S_2$, $T_2$, $S_3$ is largely due to the interaction
of two  bonding rungs separated by distances 1, 1, 2, 2, and 3 respectively,
while the formation of $T_3$ is much more complicated.
For comparison, also shown in Fig. \ref{fig_mk_even_yp1_xp2} is
the dispersion for the one-hole bonding state, and the dispersion of two-holes
sitting on the same rung\cite{oit99}: here two holes sitting on the same
rung have lower energy.

To check the correctness of our series calculations, we also performed an
exact diagonalization for
a finite lattice of 8 rungs with periodic boundary condition for this set of 
parameters. The results for $k=0,\pi$ are also 
shown in Fig. \ref{fig_mk_even_yp1_xp2}. There is remarkable agreement for the state 
$S_1$, but the
finite lattice calculation is not accurate enough to give the other bound states
sitting very close to the continuum.

\begin{figure}[htb]
  { \centering
    \includegraphics[width=\figwidth]{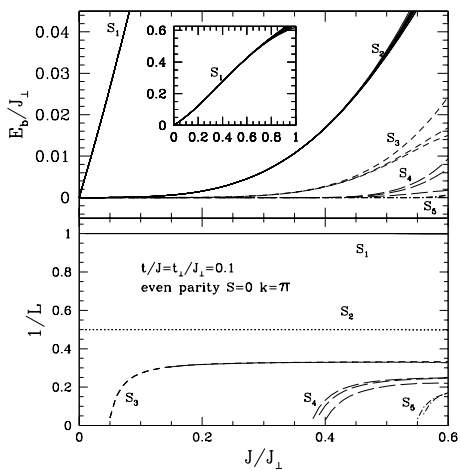}
    \caption{The binding energy $E_b$ and the inverse of the 
         coherence length $1/L$ versus  $J/J_{\perp}$ for the singlet bound states in
        the even parity channel at
          $t/J=t_{\perp}/J_{\perp}=0.1$ and $k=\pi$.
          For the binding energy of $S_1$ and $S_2$,
         the results of several different integrated differential approximants to
         the series are shown, while for other curves, the results of the three
					highest orders are plotted.
     \label{fig_Eb_L_x_ord_p1_S0_Pi_even}}  }
\end{figure}

\begin{figure}[hbt]
  { \centering
    \includegraphics[width=\figwidth]{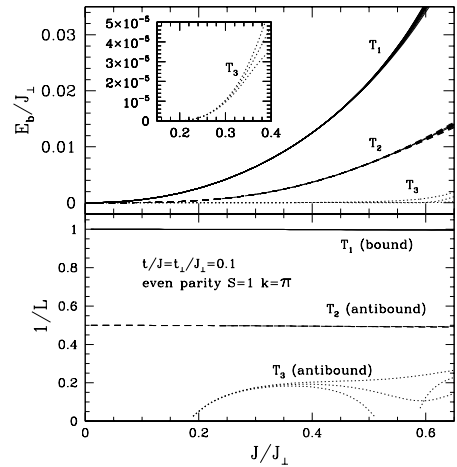}
    \caption{The binding/antibinding energy $E_b$ and the inverse of the 
         coherence length $1/L$ versus  $J/J_{\perp}$ for triplet bound/antibound states in the
                  even parity channel at
          $t/J=t_{\perp}/J_{\perp}=0.1$ and $k=\pi$.
          For the binding/antibinding energy of $T_1$ and $T_2$,
         the results of several different integrated differential approximants to
         the series are shown, while for other curves, the results of the three
					highest orders are plotted.
     \label{fig_Eb_L_x_ord_p1_S1_Pi_even}}  }
\end{figure}

\begin{figure}[htb]
  { \centering
    \includegraphics[width=\figwidth]{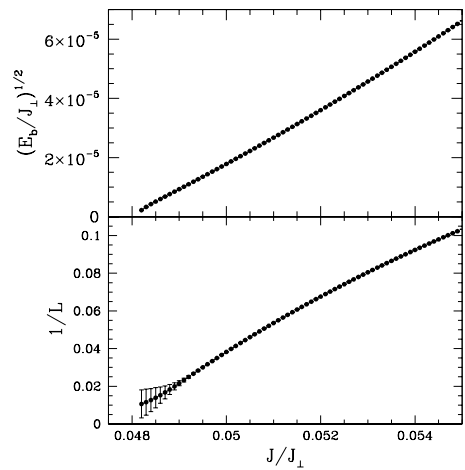}
    \caption{The square root of the binding energy $(E_b/J_{\perp})^{1/2}$ and the inverse of the 
         coherence length $1/L$ versus  $J/J_{\perp}$ for the singlet bound state $S_3$ in the
         even parity channel at
          $t/J=t_{\perp}/J_{\perp}=0.1$ and $k=\pi$.
     \label{fig_Eb_L_yp1_xc_S3_even}}  }
\end{figure}

We now study the behaviour as a function of the coupling ratio $J/J_{\perp}$,
fixing   $t/J=t_{\perp}/J_{\perp}=0.1$ and $k=\pi$.
The binding/antibinding energies $E_b$ and the inverse of the coherence length $1/L$ 
versus  $J/J_{\perp}$  are shown 
in Figs. \ref{fig_Eb_L_x_ord_p1_S0_Pi_even} and \ref{fig_Eb_L_x_ord_p1_S1_Pi_even}.
In the small $J$ limit, there are  two singlet bound states $S_1$ and $S_2$
(with coherence lengths 1 and 2, respectively), one triplet bound
state $T_1$ (coherence length 1), and one triplet antibound state
$T_2$ (coherence length 2). As $J$ increases, more and more singlet
and triplet bound states appear:  we find
that the singlet/triplet bound/antibound states $S_3$, $S_4$, $S_5$ and $T_3$
 appear at $J/J_{\perp}$ about 0.0480, 0.39, 0.55 and 0.180
respectively. To find out the behaviour near
the threshold, we plot in Fig. \ref{fig_Eb_L_yp1_xc_S3_even} the
square root of the binding energy $(E_b/J_{\perp})^{1/2}$ and the inverse of the 
coherence length $1/L$ versus  $J/J_{\perp}$ for singlet bound state $S_3$.
We can see that near threshold, the curves are nearly straight lines, which
implies that near threshold $E_b \propto (x-x_c)^2$ and $L\propto (x-x_c)^{-1}$,
corresponding to ``threshold indices"  2 and -1, respectively,
where $x=J/J_{\perp}$.
For a fixed value of $J/J_{\perp}$, the  threshold behaviour as a
function of momentum is $E_b \propto (k-k_c)^2$,
$L\propto (k-k_c)^{-1}$.
We expect similar behaviour for other bound/antibound states (but see Sect. IV).

\begin{figure}[hbt]
  { \centering
    \includegraphics[width=\figwidth]{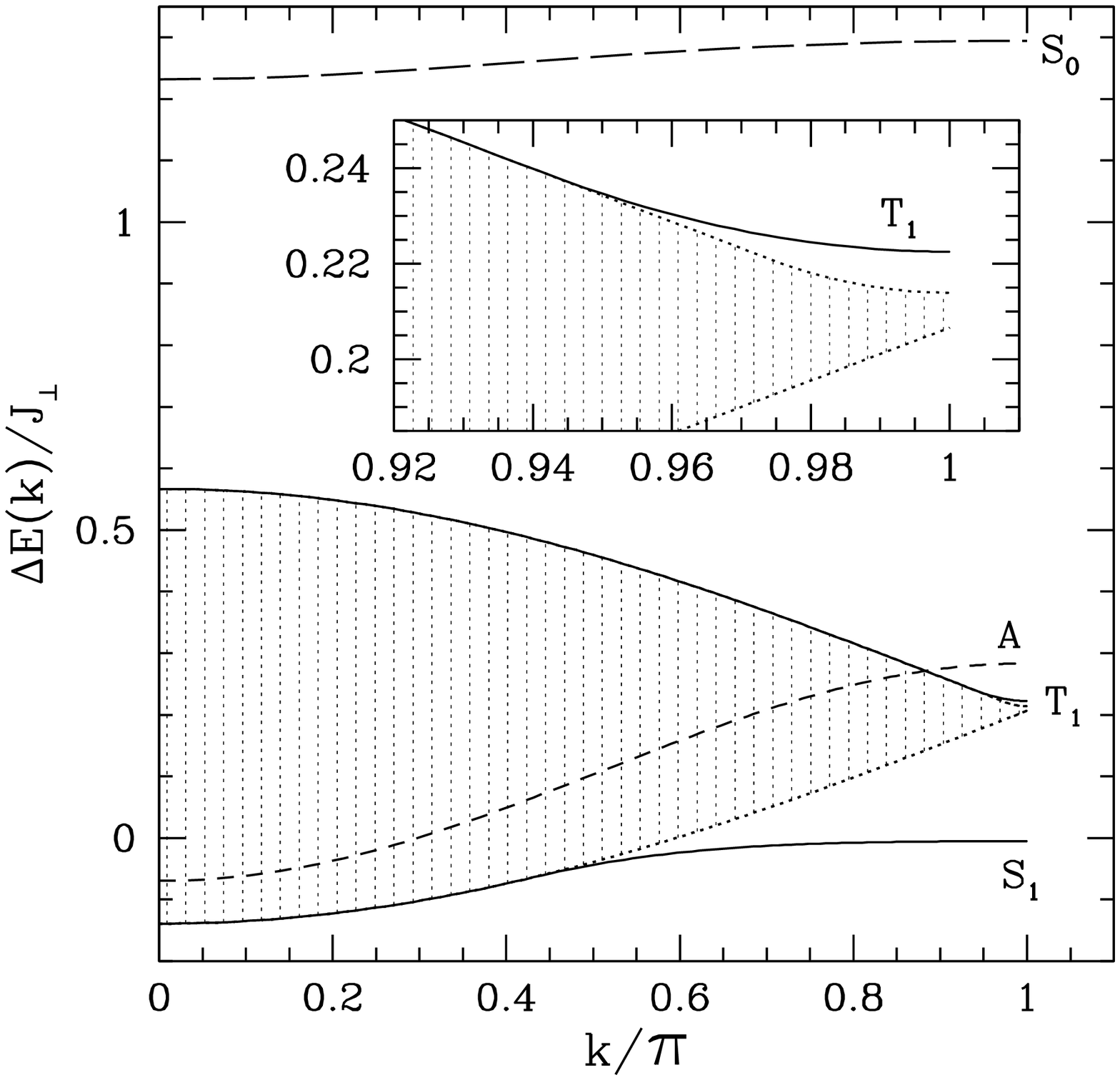}
    \caption{The excitation spectrum for the even parity channel with
          $t/J=t_{\perp}/J_{\perp}=1$ and $J/J_{\perp}=0.2$. 
          Beside the two-particle
          continuum (gray shaded), there is
          one singlet bound state ($S_1$)  below the continuum, and one
          triplet antibound state ($T_1$) above the continuum.
          The insets enlarges the region near 
          $k=\pi$ to show $T_1$ 
          above the continuum. The curve labeled  A is the dispersion of
          one hole, while the curve labeled  $S_0$ is the dispersion of
          two holes on the same rung.
     \label{fig_mk_even_y1_xp2}}  }
\end{figure}

\begin{figure}[htb]
  { \centering
    \includegraphics[width=\figwidth]{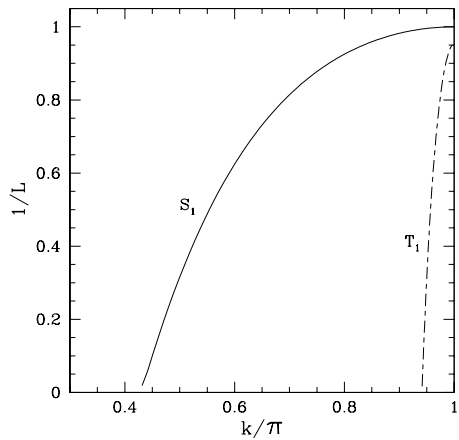}
    \caption{The inverse of the coherence length $1/L$ 
          versus momentum $k$ for the singlet $S_1$ and
          one triplet antibound state  $T_1$ in the even parity channel
          at $t/J=t_{\perp}/J_{\perp}=1$ and $J/J_{\perp}=0.2$.
     \label{fig_CL_even_y1_xp2}}  }
\end{figure}

Next we consider the case  $t/J=t_{\perp}/J_{\perp}=1$ (stronger hopping terms)
and $J/J_{\perp}=0.2$.
The two-particle excitation spectrum and the inverse of the coherence length $1/L$ 
versus momentum $k$  are shown 
in Figs. \ref{fig_mk_even_y1_xp2} and \ref{fig_CL_even_y1_xp2}.
Here there is only one singlet bound state $S_1$ below the continuum,
and one triplet antibound state ($T_1$) above it.
Both $S_1$ and $T_1$ exist only in a limited range of momenta near $k=\pi$, and
their coherence lengths $L$  at $k=\pi$ are about 1.
As comparison, also shown in Fig. \ref{fig_mk_even_y1_xp2} is
the dispersion for the one-hole bonding state, and the dispersion of two-holes
sitting on the same rung\cite{oit99}, $S_0$. In this case the state $S_0$ actually lies above
the state $S_1$, and even above the 2-particle continuum.

\begin{figure}[htb]
  { \centering
    \includegraphics[width=\figwidth]{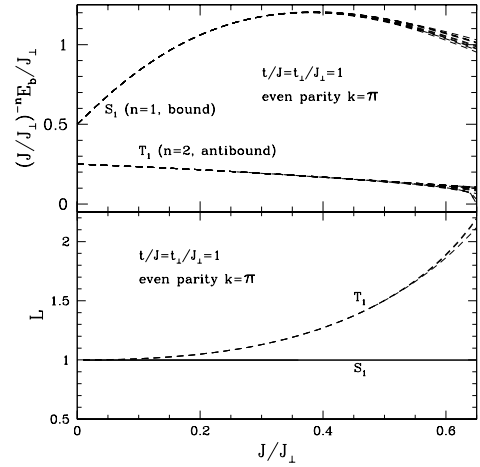}
    \caption{The (anti)-binding energy $E_b$ and the inverse of the 
         coherence length $1/L$ versus  $J/J_{\perp}$ for singlet/triplet 
         (anti)-bound states in the
         even parity channel with
          $t/J=t_{\perp}/J_{\perp}=1$ and $k=\pi$.
          For the (anti)-binding energy,
         the results of several different integrated differential approximants to
         the series are shown, while for the inverse of the 
         coherence length $1/L$, the results of the three
					highest orders are plotted.
     \label{fig_Eb_L_x_ord_y1_S01_Pi_even}}  }
\end{figure}

If we fix the values  $t/J=t_{\perp}/J_{\perp}=1$ and $k=\pi$,
the binding/antibinding energies $E_b$ and the inverse of the coherence length $1/L$ 
versus  $J/J_{\perp}$  are shown 
in Fig. \ref{fig_Eb_L_x_ord_y1_S01_Pi_even}.
One can see that 
the singlet bound state $S_1$ and the
 triplet antibound state
$T_1$ exist in the small $J$ limit, but as $J$ increases, no more new 
bound/antibound states appear.

\begin{figure}[htb]
  { \centering
    \includegraphics[width=\figwidth]{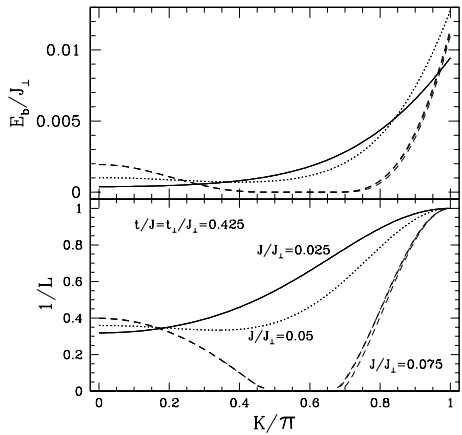}
    \caption{The binding energy $E_b$ and the inverse of the 
         coherence length $1/L$ versus $k$ for the first singlet bound state in the
         even parity channel at
          $t/J=t_{\perp}/J_{\perp}=0.425$ and $J/J_{\perp}=0.025$ (solid lines),
          0.05 (dotted lines), and 0.07 (dashed lines).  The results of the three
					highest orders are plotted. Note that for these couplings  there is a second
          singlet bound state and a triplet bound state near $k=\pi$, which are not plotted.
     \label{fig_Eb_L_k_yp425_even}}  }
\end{figure}

Sometimes a bound state can appear in the regions near both $k=0$ and $k=\pi$,
but not at intermediate momenta. 
For example for $t/J=t_{\perp}/J_{\perp}=0.425$,
the binding energy and coherence length $L$ for $J/J_{\perp}=0.025$, 0.05
and 0.075 are shown in Fig. \ref{fig_Eb_L_k_yp425_even}. 
For very small $J/J_{\perp}$ (0.025 and 0.05 in the figure),
 the singlet bound state exists over the whole range 
of momenta, but for larger $J/J_{\perp}$ (0.075 in the figure), 
 this singlet bound state only exists
 in the regions near both $k=0$ and $k=\pi$, but  not in the intermediate 
regions of momentum.

\begin{figure}[htb]
  { \centering
    \includegraphics[width=\figwidth]{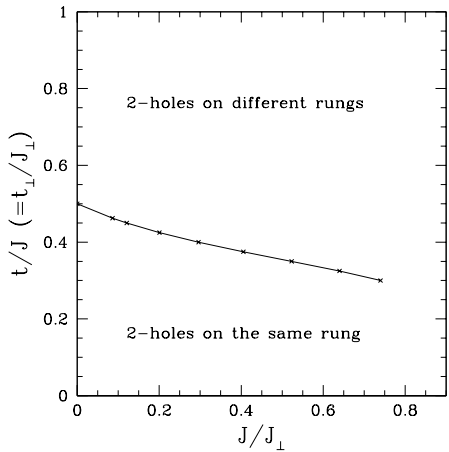}
    \caption{Boundary between different configurations of the lowest-lying 2-hole bound state.
     \label{fig_pha_2h_same_diff_rung}}  }
\end{figure}

As already seen above, the lowest energy 2-hole bound state may correspond to a pair of holes
on the same rung ($S_0$), or two holes on different rungs ($S_1$),
depending on the parameters $t/J (=t_{\perp}/J_{\perp})$ and $J/J_{\perp}$.
The boundary between these two situations is shown in Fig. \ref{fig_pha_2h_same_diff_rung}:
we will refer back to this diagram later on.


\subsection{\label{sec:oddparity}Odd-parity channel}

\begin{figure}[htb]
  { \centering
    \includegraphics[width=\figwidth]{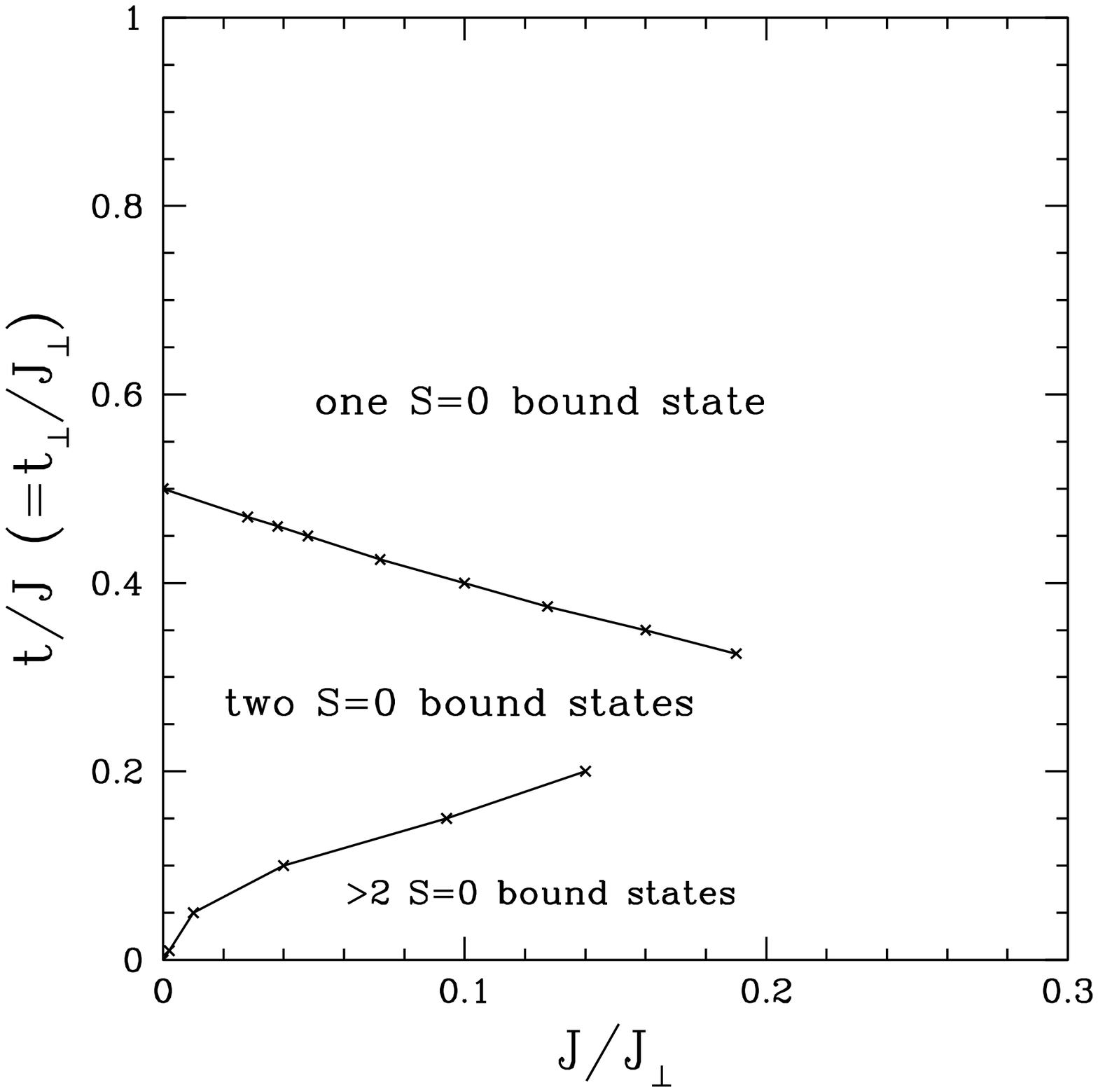}
    \caption{The number of singlet bound states in 
    the odd-parity channel in the plane of 
$t/J(=t_{\perp}/J_{\perp})$ and $J/J_{\perp}$.
     \label{fig_phadia_odd_S0}}  }
\end{figure}

\begin{figure}[hbt]
  { \centering
    \includegraphics[width=\figwidth]{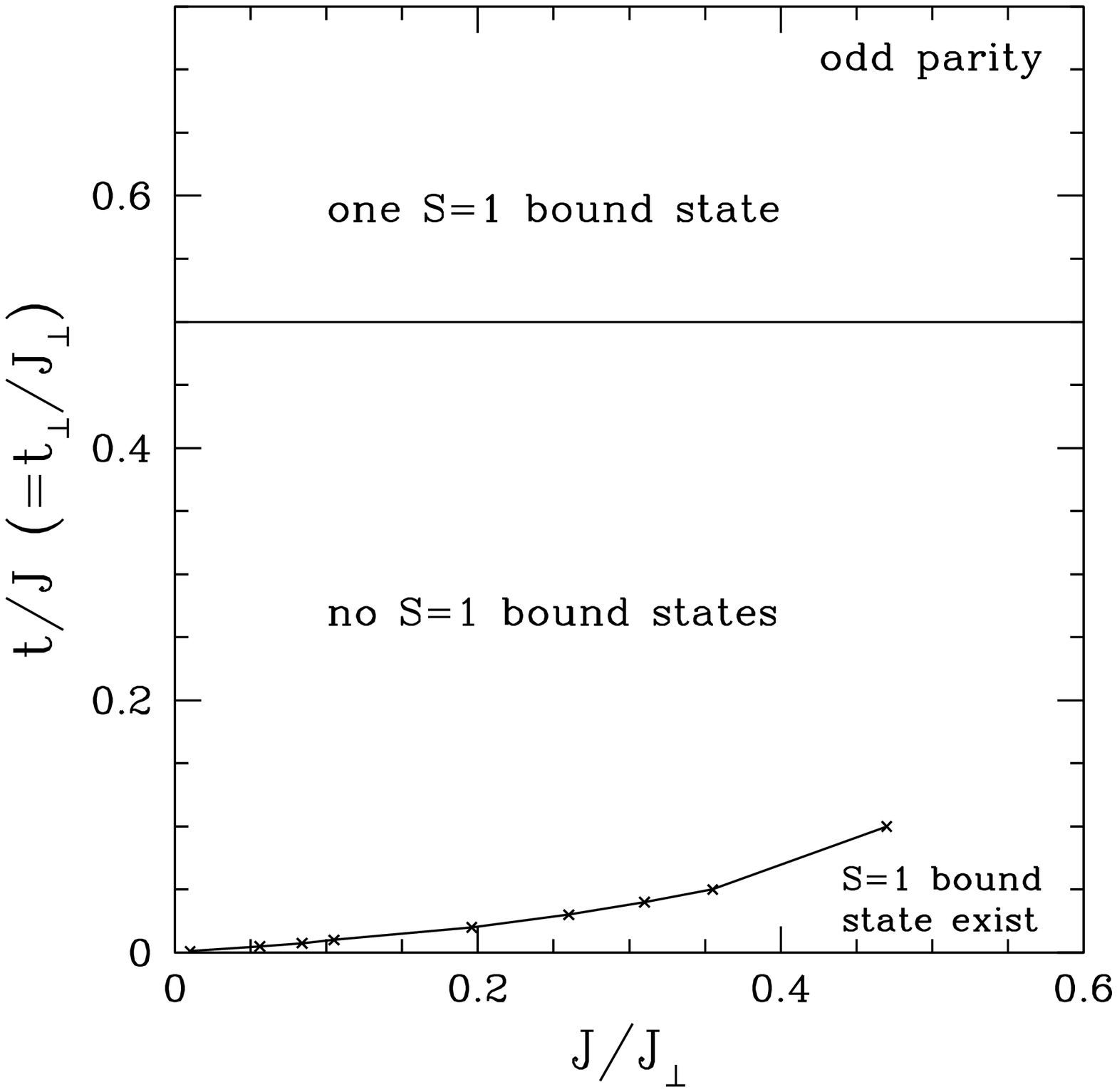}
    \caption{The number of triplet bound states in the
    odd-parity channel in the plane of 
$t/J(=t_{\perp}/J_{\perp})$ and $J/J_{\perp}$.
     \label{fig_phadia_odd_S1}}  }
\end{figure}

Among nine rung eigenstates 
for the unperturbed Hamiltonian $H_0$ in each rung, there are the following
two spin-${1\over 2}$ odd-parity electron-hole antibonding states
\bea
&&\frac{1}{\sqrt 2}(\mid 0 \downarrow \rangle - \mid \downarrow 0 \rangle )\nonumber \\
&&\frac{1}{\sqrt 2}(\mid 0 \uparrow \rangle - \mid \uparrow 0 \rangle )
\eea
The eigen energy for these two states is $t_{\perp}$.
In this section, we discuss bound states of these two 1-particle excitations. 

As for the case of even parity,  the 
number of bound states and antibound states depends
on $J/J_{\perp}$ and $t/J (=t_{\perp}/J_{\perp}$). 
Fig. \ref{fig_phadia_odd_S0} shows the diagram for the number of singlet ($S=0$)
bound states, where we can see that for large $t/J$ ($t/J \gtrsim 0.5$)
there is only one singlet bound state, while around $t/J\simeq 0.3$,
there is a region where there are two singlet bound states. For smaller $t/J$ and
$J/J_{\perp}>0$, there is a region where there are more than two singlet bound states,
and in this region, the number of bound states increases as  $J/J_{\perp}$ increases.

Fig. \ref{fig_phadia_odd_S1} shows the diagram for the number of triplet ($S=1$)
bound states. For $t/J \gtrsim 0.5$,
there is one triplet bound state, while elsewhere
no triplet bound state exists, except  in a region at  very small $t/J$.

\begin{figure}[htb]
  { \centering
    \includegraphics[width=\figwidth]{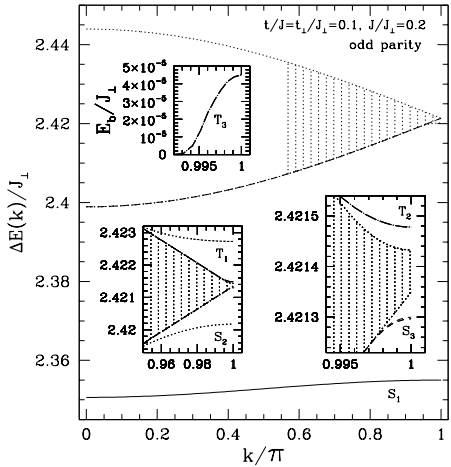}
    \caption{The excitation spectrum in the odd parity channel at
          $t/J=t_{\perp}/J_{\perp}=0.1$ and $J/J_{\perp}=0.2$. 
          Beside the two-particle
          continuum (gray shaded), there are
          three singlet bound states ($S_1$, $S_2$ and $S_3$)  below the continuum, and three
          triplet antibound states ($T_1$, $T_2$ and $T_3$) above the continuum.
          Two lower insets enlarge the region near 
          $k=\pi$ to show $S_2$, $S_3$, $T_1$ and $T_2$ 
          below/above the continuum,
          but the antibound state
          $T_3$ is still indistinguishable. To see  $T_3$,
          we plot the  antibinding energy
          versus $k/\pi$, as the top inset.  
     \label{fig_mk_odd_yp1_xp2}}  }
\end{figure}

\begin{figure}[htb]
  { \centering
    \includegraphics[width=\figwidth]{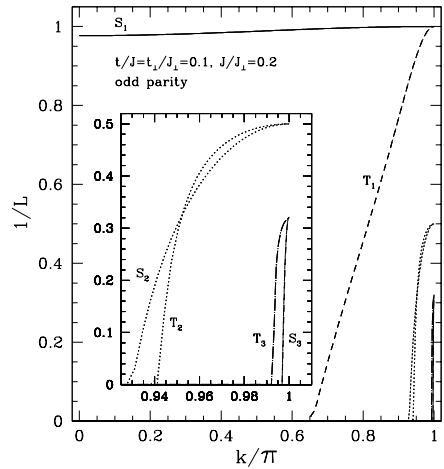}
    \caption{The inverse of the coherence length $1/L$ 
          versus momentum $k$ for three singlet ($S_1$, $S_2$ and $S_3$), 
           and three triplet ($T_1$, $T_2$ and $T_3$) 
          antibound states  for the odd parity channel
          at $t/J=t_{\perp}/J_{\perp}=0.1$ and $J/J_{\perp}=0.2$. 
          The inset enlarges the region near 
          $k=\pi$.
     \label{fig_CL_odd_yp1_xp2}}  }
\end{figure}

If we take the values $t/J=t_{\perp}/J_{\perp}=0.1$ and $J/J_{\perp}=0.2$,
the two-particle excitation spectrum and the inverse of the coherence length $1/L$ 
versus momentum $k$  are shown 
in Figs. \ref{fig_mk_odd_yp1_xp2} and \ref{fig_CL_odd_yp1_xp2}.
There are three singlet bound states ($S_1$, $S_2$ and $S_3$) 
below the continuum, and three
triplet antibound states ($T_1$, $T_2$ and $T_3$) above the continuum.
The singlet bound state $S_1$ exists for the whole range 
of momenta (its coherence length $L$ is finite also for the whole range 
of momenta), while other
bound/antibound states exist only in a limited range of momenta near $k=\pi$. 
At $k=\pi$, the coherence length $L$ for $S_1$ and $T_1$ is 1,  for
$S_2$ and $T_2$ it is about 2, while $S_3$ and $T_3$ have coherence length $L$ about 3.
This means that the formation of bound/antibound
 states  $S_1$, $T_1$ , $S_2$, $T_2$, $S_3$ and $T_3$ is largely due to the interaction
of two  antibonding rungs separated by distances 1, 1, 2, 2, 3, and 3 respectively.

\begin{figure}[htb]
  { \centering
    \includegraphics[width=\figwidth]{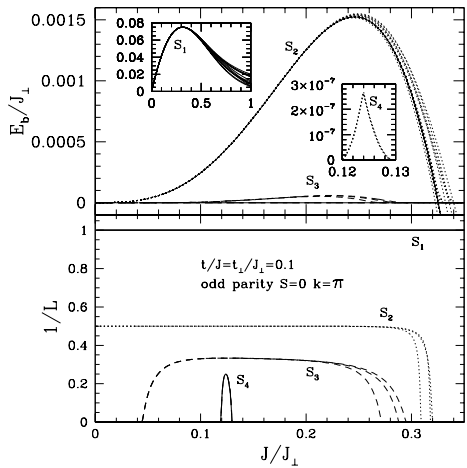}
    \caption{The binding energy $E_b$ and the inverse of the 
         coherence length $1/L$ versus  $J/J_{\perp}$ for singlet bound states in the
         odd parity channel at
          $t/J=t_{\perp}/J_{\perp}=0.1$ and $k=\pi$.
          For the binding energy of $S_1$ and $S_2$,
         the results of several different integrated differential approximants to
         the series are shown, while for other curves, the results of the three
					highest orders are plotted.
     \label{fig_Eb_L_x_ord_p1_S0_Pi_odd}}  }
\end{figure}

\begin{figure}[htb]
\vspace{1cm}
  { \centering
    \includegraphics[width=\figwidth]{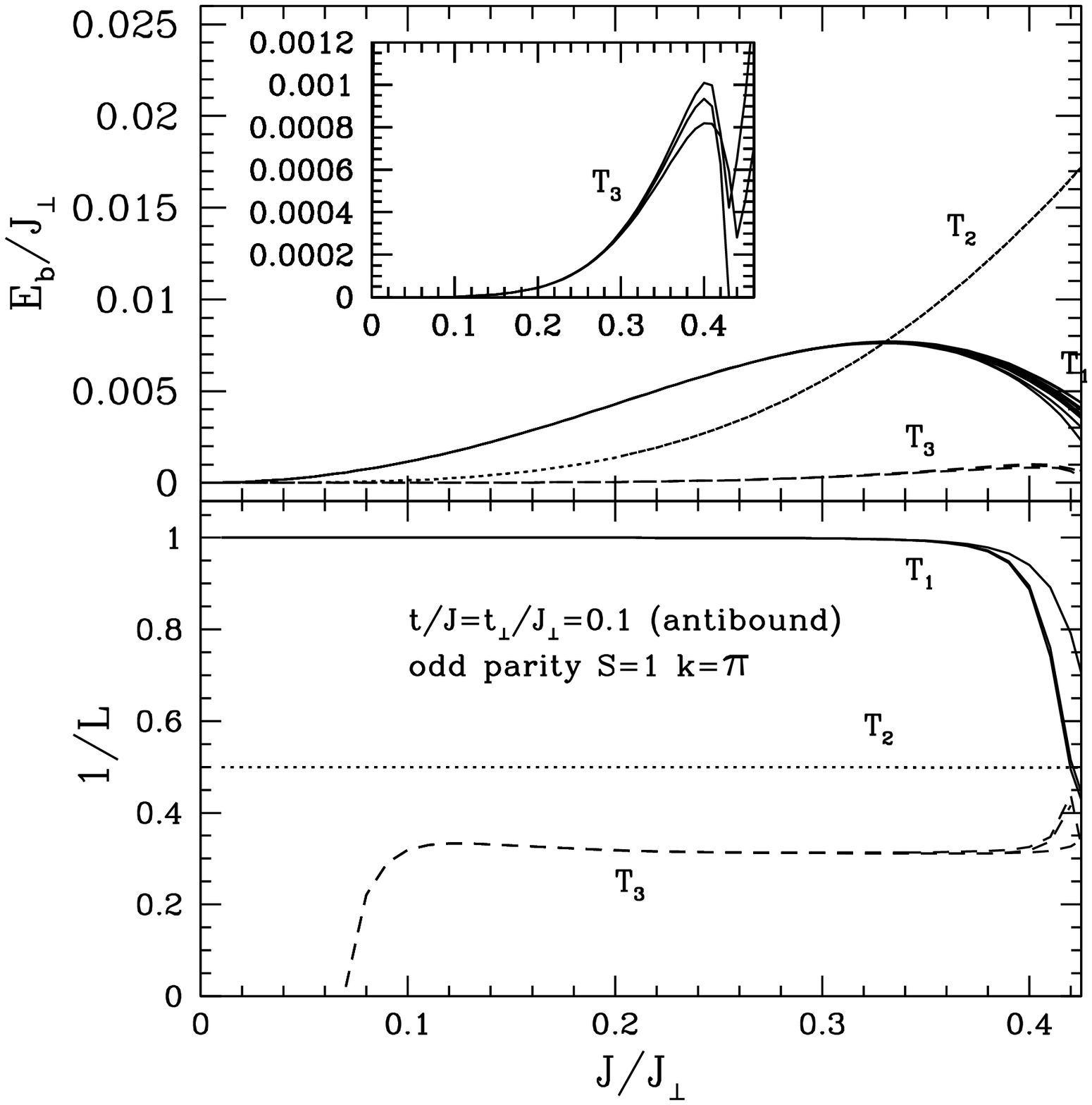}
    \caption{The antibinding energy $E_b$ and the inverse of the 
         coherence length $1/L$ versus  $J/J_{\perp}$ for triplet antibound states in the
         odd parity channel with
          $t/J=t_{\perp}/J_{\perp}=0.1$ and $k=\pi$.
          For the antibinding energy of $T_1$ and $T_2$,
         the results of several different integrated differential approximants to
         the series are shown, while for other curves, the results of the three
					highest orders are plotted.
     \label{fig_Eb_L_x_ord_p1_S1_Pi_odd}}  }
\end{figure}

If we fix the values  $t/J=t_{\perp}/J_{\perp}=0.1$ and $k=\pi$,
the binding/antibinding energies $E_b$ and the inverse of the coherence length $1/L$ 
versus  $J/J_{\perp}$  are shown 
in Figs. \ref{fig_Eb_L_x_ord_p1_S0_Pi_odd} and \ref{fig_Eb_L_x_ord_p1_S1_Pi_odd}.
One can see that at the small $J$ limit, we have two singlet bound states $S_1$ and $S_2$
(with coherence lengths 1 and 2, respectively),  and two triplet antibound states
$T_1$ and $T_2$ (with coherence lengths 1 and 2, respectively). As $J$ increases, more and more singlet
bound state and triplet antibound states turn up, and in our calculations, we find
that states $S_3$, $S_4$, and $T_3$
 appear at $J/J_{\perp}$ about 0.05, 0.12, and 0.07 
respectively, while 
$S_2$, $S_3$, $S_4$, and $T_1$ disappear at $J/J_{\perp}$ about 0.32, 0.29, 0.13 and 0.46,
respectively. Thus state $S_4$ pops out below the continuum only very briefly.
The threshold index for the binding energy of
$S_2$ and $T_1$ is about 1, rather than 2, as found by Dlog Pad\'e approximants to the series.

\begin{figure}[htb]
  { \centering
    \includegraphics[width=\figwidth]{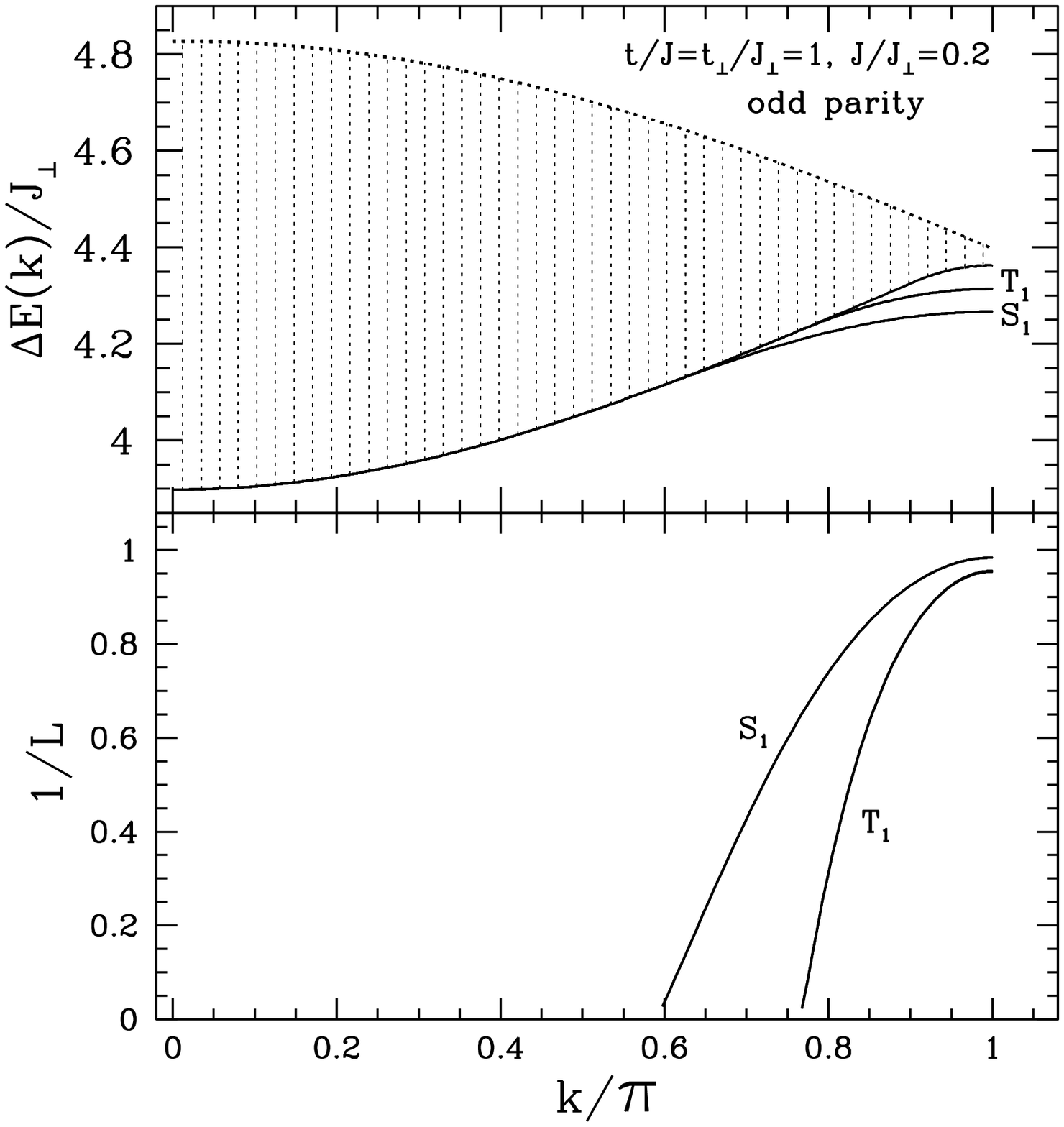}
    \caption{The excitation spectrum and the inverse of the 
         coherence length $1/L$ in the odd parity channel at
          $t/J=t_{\perp}/J_{\perp}=1$ and $J/J_{\perp}=0.2$. 
          Beside the two-particle
          continuum (gray shaded), there are
          one singlet  ($S_1$) and one
          triplet bound state ($T_1$) below the continuum.
     \label{fig_mk_odd_y1_xp2}}  }
\end{figure}

For larger hopping terms  $t/J=t_{\perp}/J_{\perp}=1$ and $J/J_{\perp}=0.2$,
the two-particle excitation spectrum and the inverse of the coherence length $1/L$ 
versus momentum $k$  are shown 
in Fig. \ref{fig_mk_odd_y1_xp2}.
Now  there is  one singlet bound state $S_1$
and one triplet bound state ($T_1$) below the continuum,  but no antibound state.
Both $S_1$ and $T_1$ exist only in a limited range of momenta near $k=\pi$, and
their coherence lengths $L$  at $k=\pi$ are about 1.

\begin{figure}[htb]
  { \centering
    \includegraphics[width=\figwidth]{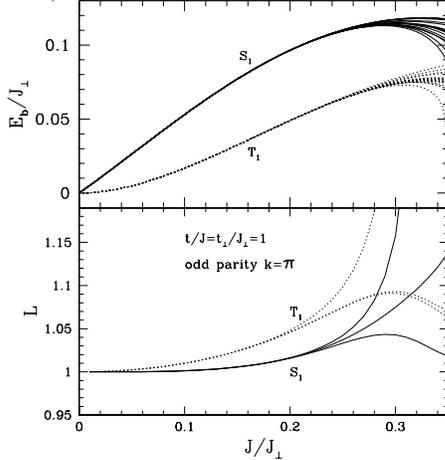}
    \caption{The binding energy $E_b$ and the inverse of the 
         coherence length $1/L$ versus  $J/J_{\perp}$ for singlet and triplet bound state 
         in the odd parity channel at
          $t/J=t_{\perp}/J_{\perp}=1$ and $k=\pi$.
          For the binding energy,
         the results of several different integrated differential approximants to
         the series are shown, while for the coherence length, the results of the three
					highest orders are plotted.
     \label{fig_Eb_L_x_ord_y1_S01_Pi_odd}}  }
\end{figure}

If we fix the values  $t/J=t_{\perp}/J_{\perp}=1$ and $k=\pi$,
the binding energies $E_b$ and the inverse of the coherence length $1/L$ 
versus  $J/J_{\perp}$  are shown 
in Fig. \ref{fig_Eb_L_x_ord_y1_S01_Pi_odd}.
The singlet bound state $S_1$ and the
 triplet bound state
$T_1$ exist as long as $J\not=0$ , but  no more new 
bound/antibound states appear as $J$ increases.

\section{\label{sec:4holes}Four-hole bound states and phase separation}

\begin{figure}[htb]
  { \centering
    \includegraphics[width=\figwidth]{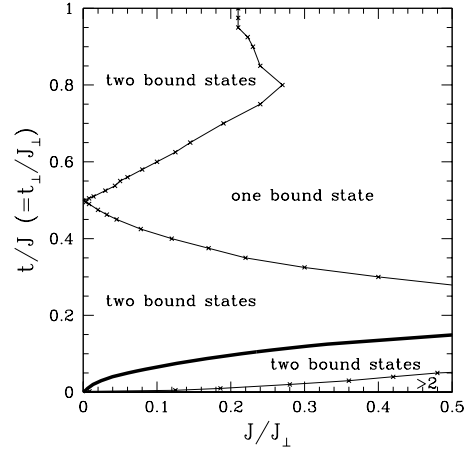}
    \caption{The number of singlet 4-hole bound states  in the plane of 
$t/J(=t_{\perp}/J_{\perp})$ and $J/J_{\perp}$. 
Note that the bold solid line is the region where
there is only one bound state, whereas $>2$ is the region that there are more than 2 bound states.
     \label{fig_phadia_4holes}}  }
\end{figure}

Among nine rung eigenstates 
for the unperturbed Hamiltonian $H_0$ on each rung, there is one state of a two-hole pair with
eigenenergy 0. 
In this section, we discuss bound states of 2 two-hole pair excitations, corresponding to the half filled
system doped with 4 holes.

As for the case of 2-hole bound states,  the 
number of bound states and antibound states depends
on $J/J_{\perp}$ and $t/J (=t_{\perp}/J_{\perp}$). 
Fig. \ref{fig_phadia_4holes} shows the  number of singlet ($S=0$)
bound states in various regions of parameter space.
For large $t/J$ ($t/J \gtrsim 0.3$) and large $J/J_{\perp}$,
there is only one singlet bound state, while elsewhere
 there are two singlet bound states, except for a peculiar very narrow region (given by the
 bold solid line) where there is only one bound state, and a small region at low $t/J$ (labelled
 by $>2$) where there are more than two bound states.

\begin{figure}[htb]
  { \centering
    \includegraphics[width=\figwidth]{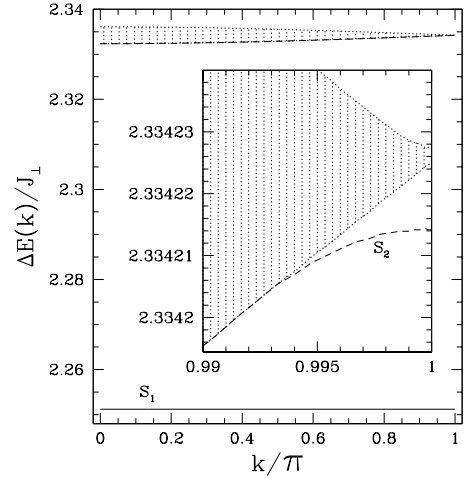}
    \caption{The excitation spectrum  for 4-hole bound states at
          $t/J=t_{\perp}/J_{\perp}=0.1$ and $J/J_{\perp}=0.15$. 
          Beside the two-particle
          continuum (gray shaded), there are
          two singlet bound states ($S_1$ and $S_2$)  below the continuum.
          The inset enlarges the region near 
          $k=\pi$ to show $S_2$ 
          below the continuum.
     \label{fig_mk_4hole_yp1_xp15}}  }
\end{figure}

\begin{figure}[htb]
  { \centering
    \includegraphics[width=\figwidth]{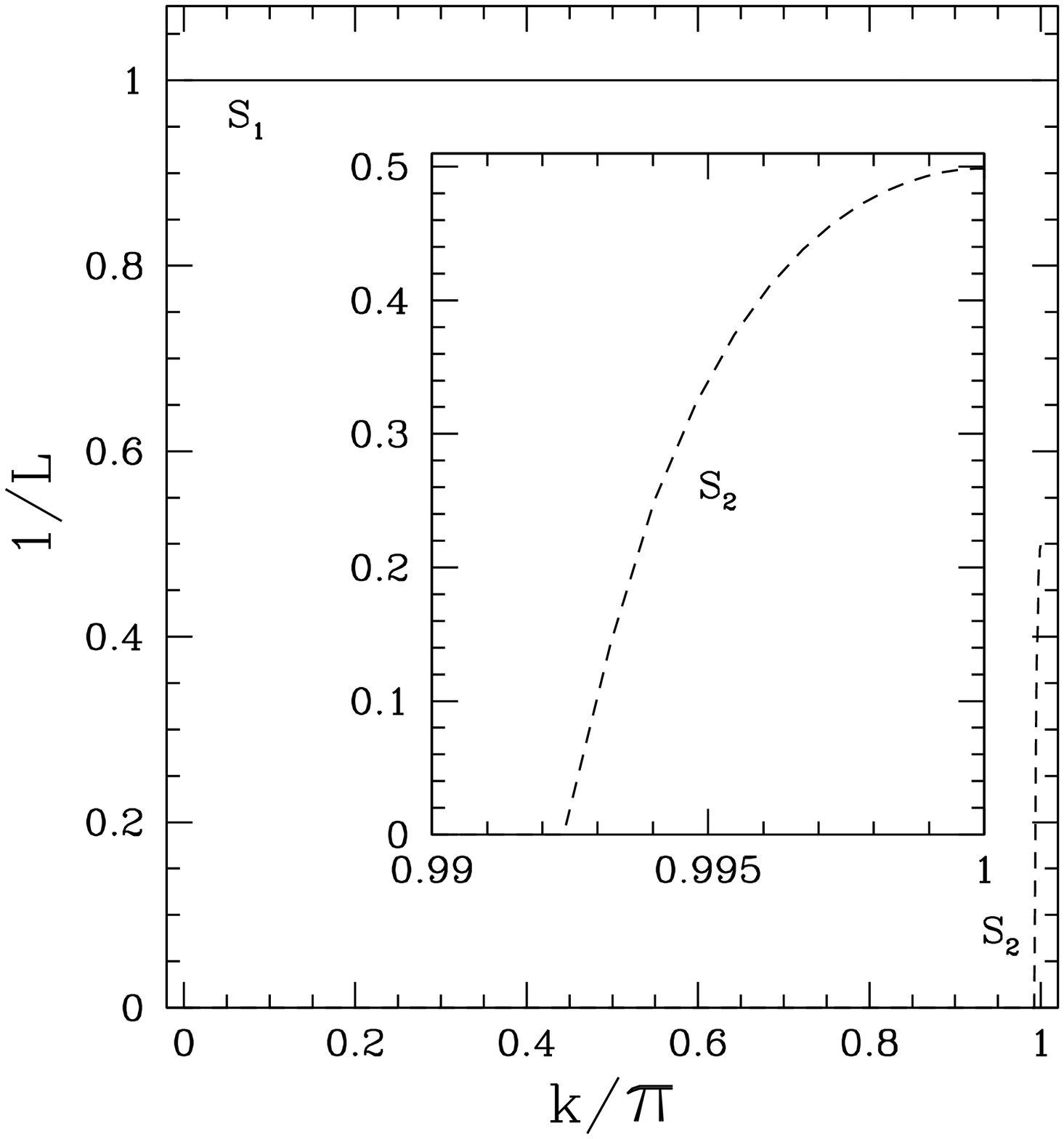}
    \caption{The inverse of the coherence length $1/L$ 
          versus momentum $k$ for two singlet ($S_1$ and $S_2$) 
          4-hole bound states 
          at $t/J=t_{\perp}/J_{\perp}=0.1$ and $J/J_{\perp}=0.15$. 
          The inset enlarges the region near 
          $k=\pi$.
     \label{fig_CL_4holes_yp1_xp15}}  }
\end{figure}

If we take the values $t/J=t_{\perp}/J_{\perp}=0.1$ and $J/J_{\perp}=0.15$
(we take $J/J_{\perp}=0.15$, rather than $J/J_{\perp}=0.2$ as before, because
 at $J/J_{\perp}=0.2$ we have only one bound state),
the two-particle excitation spectrum and the inverse of the coherence length $1/L$ 
versus momentum $k$  are shown 
in Figs. \ref{fig_mk_4hole_yp1_xp15} and \ref{fig_CL_4holes_yp1_xp15}.
There are two singlet bound states ($S_1$ and $S_2$) 
below the continuum.
The singlet bound state $S_1$ exists for the whole range 
of momenta (its coherence length $L$ is finite also for the whole range 
of momenta), while bound  state $S_2$ exists only in a limited range of momenta near $k=\pi$. 
The coherence length $L$ for $S_1$  is 1, while the coherence length $L$ for
$S_2$ at $k=\pi$ is about 2.
This means that the formation of bound
 states  $S_1$ and $S_2$ is largely due to the interaction
of two hole-pair  rungs separated by 0 and 1 singlet rung, respectively.

\begin{figure}[htb]
  { \centering
    \includegraphics[width=\figwidth]{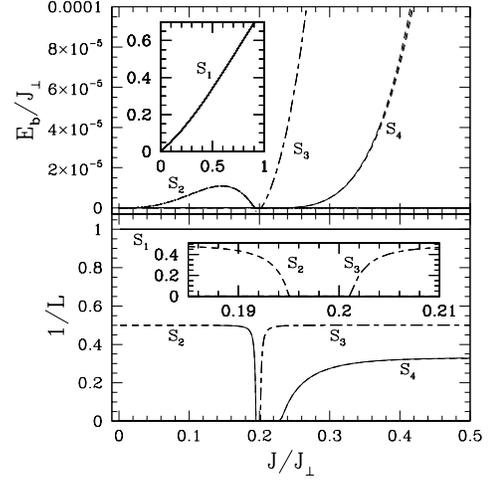}
    \caption{The binding energy $E_b$ and the inverse of the 
         coherence length $1/L$ versus  $J/J_{\perp}$ for singlet 4-hole bound/antibound states  at
          $t/J=t_{\perp}/J_{\perp}=0.1$ and $k=\pi$.
          For the binding energy of $S_1$ and $S_2$,
         the results of several different integrated differential approximants to
         the series are shown, while for other curves, the results of the three
					highest orders are plotted.
     \label{fig_Eb_L_x_ord_p1_Pi_4holes}}  }
\end{figure}

If we fix the values  $t/J=t_{\perp}/J_{\perp}=0.1$ and $k=\pi$,
the binding/antibinding energies $E_b$ and the inverse of the coherence length $1/L$ 
versus  $J/J_{\perp}$  are shown 
in Fig. \ref{fig_Eb_L_x_ord_p1_Pi_4holes}.
One can see that in the small $J$ limit, we have two singlet bound states $S_1$ and $S_2$
(with coherence length being 1 and 2, respectively). As $J$ increases, $S_2$ disappears at
$J/J_{\perp}\simeq 0.193$, while a new antibound state $S_3$ and a new bound state $S_4$
appear at $J/J_{\perp}\simeq 0.205$ and 0.235, (coherence length at larger $J$
being 2 and 3), respectively.
This interval corresponds to the peculiar region given by the bold line in 
Fig. \ref{fig_phadia_4holes}. The threshold index as the binding energy of
$S_2$ disappears appears to be about 1 , rather than 2. 
Estimates for the threshold point and index from the $[n/m]$
Dlog Pad\'e approximants to the series for the binding energy are given in Table \ref{tab3}.
We cannot calculate a series directly for the coherence length, but from our numerical
calculations, the threshold behaviour  for the coherence length of $S_2$ is also unlikely to be
the standard $L\propto (x-x_c)^{-1}$, but seems to be a weaker divergence.
The dynamical reason for this unusual threshold behaviour is unclear\cite{olegcomment}.
Note that even in this case, the threshold behaviour as a
function of momentum still retains the form: $E_b \propto (k-k_c)^2$,
$L\propto (k-k_c)^{-1}$.

\begin{figure}[htb]
  { \centering
    \includegraphics[width=\figwidth]{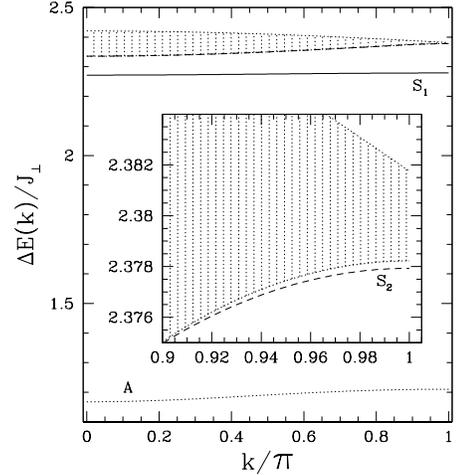}
    \caption{The excitation spectrum for 4-holes bound state at
          $t/J=t_{\perp}/J_{\perp}=1$ and $J/J_{\perp}=0.15$. 
          Beside the two-particle
          continuum (gray shaded), there are
          two singlet  ($S_1$ and $S_2$) below the continuum.
          Curve  A is the dispersion of the two-holes sitting in the same rung. 
     \label{fig_mk_4holes_y1_xp15}}  }
\end{figure}

\begin{figure}[htb]
  { \centering
    \includegraphics[width=\figwidth]{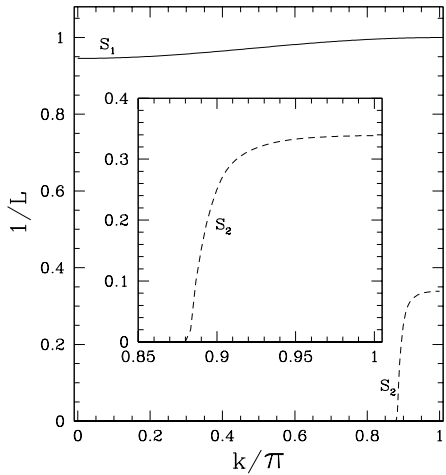}
    \caption{The inverse of the coherence length $1/L$ 
          versus momentum $k$ for two singlet ($S_1$ and $S_2$) 
          4-holes bound states 
          at $t/J=t_{\perp}/J_{\perp}=1$ and $J/J_{\perp}=0.15$. 
          The inset enlarges the region near 
          $k=\pi$.
     \label{fig_CL_4holes_y1_xp15}}  }
\end{figure}

For larger hopping terms  $t/J=t_{\perp}/J_{\perp}=1$ and $J/J_{\perp}=0.15$,
the two-particle excitation spectrum and the inverse of the coherence length $1/L$ 
versus momentum $k$  are shown 
in Figs. \ref{fig_mk_4holes_y1_xp15} and \ref{fig_CL_4holes_y1_xp15}.
Again there are  two singlet bound states $S_1$ and $S_2$
 below the continuum,  but no antibound state.
The singlet bound state $S_1$ exists for the whole range 
of momenta (its coherence length $L$ is about 1 also for the whole range 
of momenta), while bound  state $S_2$ exists only in a limited range of momenta near $k=\pi$
(its coherence length $L$ is about 2 at $k=\pi$). 

\begin{figure}[htb]
  { \centering
    \includegraphics[width=\figwidth]{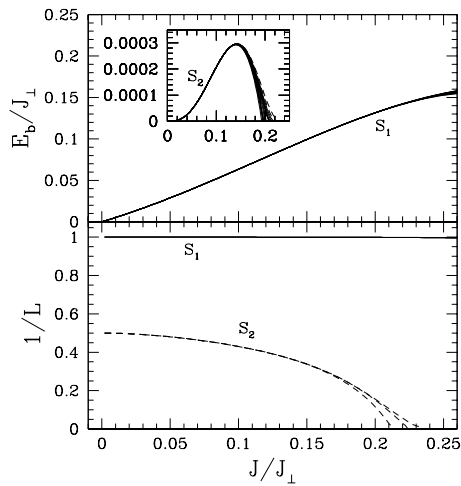}
    \caption{The binding energy $E_b$ and the inverse of the 
         coherence length $1/L$ versus  $J/J_{\perp}$ for two singlet 4-holes bound states 
          at   $t/J=t_{\perp}/J_{\perp}=1$ and $k=\pi$.
          For the binding energy,
         the results of several different integrated differential approximants to
         the series are shown, while for the coherence length, the results of the three
					highest orders are plotted.
     \label{fig_Eb_L_x_ord_y1_Pi_4holes}}  }
\end{figure}

If we fix the values  $t/J=t_{\perp}/J_{\perp}=1$ and $k=\pi$,
the binding energies $E_b$ and the inverse of the coherence length $1/L$ 
versus  $J/J_{\perp}$  are shown 
in Fig. \ref{fig_Eb_L_x_ord_y1_Pi_4holes}.
The two singlet bound states $S_1$ and $S_2$ exist as long as $J\not=0$ , 
but as $J$ increases, no more new 
bound/antibound states appear, but $S_2$ disappears at $J/J_{\perp}\simeq 0.22$.

The compressibility $\kappa$ is given by
\be
\kappa^{-1} = \rho^2 {\partial^2 \epsilon \over \partial \rho^2}
\ee
where $\epsilon (\rho)$ is the energy density of the ladder with hole density $\rho$. 
For a finite system at half-filling this can be replaced by the discrete version
\bea
\kappa^{-1} &=& (L/2) [ E_0 (\frac{1}{2} {\rm ~filled}) - 2 E_0 ({\rm 2~holes}) + E_0 ({\rm 4~holes}) ] \nonumber \\
 &=& (L/2) [ \Delta E_0 ({\rm 4~holes}) - 2 \Delta E_0 ({\rm 2~holes})  ]    \label{eq_kappa}
\eea
where for a ladder of $L$ rungs, $\Delta E_0$ is the minimum energy gap for 2 or 4 holes, and
this minimum energy gap is at momentum $k=0$. 
Now if the 2-hole pair on a single rung is the lowest-energy state in the 2-hole
sector, and at momentum $k=0$ two 2-hole rungs form a bound state in the 4-hole sector, then it immediately
follows that $\kappa^{-1}$ as defined by Eq. \ref{eq_kappa}
diverges to negative infinity as $L\to \infty$, signalling {\it phase separation}.
Comparing Figures \ref{fig_phadia_4holes} and \ref{fig_pha_2h_same_diff_rung},
we see that these conditions apply, and phase separation occurs, over the
whole of the lower region of Figure \ref{fig_pha_2h_same_diff_rung}.

Outside this region, we know the energy of the lowest state in the 2-hole sector,
consisting of a bound pair of holes on different rungs (denoted $S_1$, previously).
The bound state of two 2-hole rungs may not be the lowest-energy state in the
4-hole sector; but at least it allows us to place an upper limit on $\kappa^{-1}$
as defined by Eq. \ref{eq_kappa}. We find that this limit is negative, and therefore
phase separation must occur, for $t/J (=t_{\perp}/J_{\perp})$ anywhere below a value about 0.5, 
independent of $J/J_{\perp}$ - although due to numerical uncertainty, this
estimate cannot be made very precise.  This is in good agreement with previous
estimates: Rommer {\it et al.}\cite{rom00}, for instance, find the phase
separation boundary  at half-filling to lie at
$J/t = 2.156(2)$, or $t/J=0.464$, from a density matrix renormalization group calculation for
the isotropic ladder.


\section{\label{sec:con}Discussion and Summary}

We have used linked-cluster series expansion methods\cite{tre00} to study
2-hole bound states in the $t-J$ ladder. The series only converge well for 
$J/J_{\perp}\lesssim 0.5$, so we are unable to say very much about the isotropic case
$J=J_{\perp}$.
Dispersion relations and coherence lengths have been computed for various bound states in
even- and odd- parity channels.
A rich spectrum of these states has emerged.

The most striking feature of the results is the appearance of multiple bound
states in both singlet $S=0$ and triplet $S=1$ channels as the ratio 
$t/J (=t_{\perp}/J_{\perp})$ goes to zero for finite $J/J_{\perp}$.
What is the cause of this phenomenon? 
The limit $t/J \to 0$ corresponds to the antiferromagnetic Heisenberg ladder
with 2 static holes. Beyond states $S_0$ and $S_1$, all the multiple bound states
($S_2, \cdots$) only emerge near $k=\pi$, and all have small binding energies:
their dynamical significance is unclear.
%

Another interesting feature is that in the even-parity channel, at least, all
the triplet $S=1$ bound states 
seem to disappear as the isotropic limit $J/J_{\perp}\to 1$
is approached, leaving only singlet $S=0$ bound states. 
Other studies\cite{jur01} have found the same thing.
We have no 
explanation for this phenomenon, either.

Jurecka and Brenig\cite{jur01} have put forward a semi-analytic theory of 2-hole
excitations in the $t-J$ ladder, based on the dimerized rung picture, with
elementary excitations consisting of the single-hole pseudofermion
excitations and the 2-hole singlet excitation on a single rung, plus singlet
and triplet ``bond boson" operators for spin excitations, and applying a linearized
Holstein-Primakoff approach. They present a spectrum showing only one
$S=0$ and one $S=1$ bound state, at couplings which are not accessible
to our series approach.

The question of phase separation has also been explored. We have been able to calculate
the energy of a 4-hole state consisting of two 2-hole rungs, giving an
upper limit on the inverse  compressibility at half-filling.
Hence it can be shown that phase separation occurs for $t/J\lesssim 0.5$,
in agreement with other studies\cite{rom00}.
Most of the interesting spectral features occur, unfortunately, in the phase
separated region.

We have also explored the threshold behaviour of these bound/antibound states as
they emerge from the continuum.
As a function of momentum, the binding energy near threshold behaves as
$E_b \propto (k-k_c)^2$, and the coherence length as  $L\propto (k-k_c)^{-1}$
in every case we have studied so far. This appears to be standard:
for instance, a model of the bound states of two magnons by Sushkov and Kotov\cite{sus98}
gives this behaviour. As a function of coupling $x=J/J_{\perp}$,
the typical threshold behaviour is similar,
$E_b \propto (x-x_c)^2$,  $L\propto (x-x_c)^{-1}$. We have found examples,
however, where the binding energy appears to behave as
$E_b \propto (x-x_c)$, see Sect. IIIB and IV.

In conclusion, then, we have discovered a rich and diverse spectrum of 2-hole
bound states in the $t-J$ ladder at half-filling. 
They exhibit a number of interesting feature 
which call for a theoretical explanation; and fitting these data will provide
a stringent test of any theoretical model of the system. It would also be interesting
if it were possible to test these features experimentally.

\begin{acknowledgments}
We are grateful to Prof. Jaan Oitmaa and Oleg Sushkov for useful discussions.
This work  forms part of 
a research project supported by a grant
from the Australian Research Council.
The computations have been performed on an AlphaServer SC
 computer. We are grateful for the computing resources provided
 by the Australian Partnership for Advanced Computing (APAC)
National Facility.
\end{acknowledgments}


\newpage
\bibliography{basename of .bib file}

\begin{references}
\bibitem{olegcomment}Prof. O. Sushkov has suggested that it may correspond to a
``d-wave" rather than ``s-wave" threshold.
\end{references}


\begin{table*}
\caption{The nine rung states and their energies at $t=J=0$.}\label{tab1}
\begin{ruledtabular}
\begin{tabular}{|c|c|c|c|}
\multicolumn{1}{|l|}{No.} &\multicolumn{1}{l|}{Eigenstate}
&\multicolumn{1}{l|}{Eigenvalue} &\multicolumn{1}{l|}{Name} \\
\hline
1  &  $\frac{1}{\sqrt 2}(\mid \uparrow \downarrow \rangle - \mid \downarrow \uparrow \rangle )$ &    $-J_{\perp}$  &  singlet  \\
\hline
2  & $\mid \downarrow \downarrow \rangle  $  &    $0$  &  triplet ($S^z_{\rm tot}=-1$)  \\
\hline
3  & $\frac{1}{\sqrt 2}(\mid \uparrow \downarrow \rangle + \mid \downarrow \uparrow \rangle )$ & $0$  &  triplet ($S^z_{\rm tot}=0$)  \\
\hline
4  & $\mid \uparrow \uparrow \rangle  $  &    $0$  &  triplet ($S^z_{\rm tot}=1$)  \\
\hline
5  & $\mid 00 \rangle$  &    $0$  &  hole-pair singlet  \\
\hline
6  & $\frac{1}{\sqrt 2}(\mid 0 \downarrow \rangle + \mid \downarrow 0 \rangle )$ & $-t_{\perp}$  & electron-hole bonding ($S^z_{\rm tot}=-\frac{1}{2}$)  \\
\hline
7  & $\frac{1}{\sqrt 2}(\mid 0 \uparrow \rangle + \mid \uparrow 0 \rangle )$ & $-t_{\perp}$  & electron-hole bonding ($S^z_{\rm tot}=\frac{1}{2}$)  \\
\hline
8  & $\frac{1}{\sqrt 2}(\mid 0 \downarrow \rangle - \mid \downarrow 0 \rangle )$ & $t_{\perp}$  & electron-hole antibonding ($S^z_{\rm tot}=-\frac{1}{2}$)  \\
\hline
9  & $\frac{1}{\sqrt 2}(\mid 0 \uparrow \rangle - \mid \uparrow 0 \rangle )$ & $t_{\perp}$  & electron-hole antibonding ($S^z_{\rm tot}=\frac{1}{2}$)  \\
\end{tabular}
\end{ruledtabular}
\end{table*}

\begin{table*}
\squeezetable
\caption{Series coefficients for dimer expansions for the energy gap $E/J_{\perp}$ 
of two singlet bound states ($S_1$ and $S_2$), 
          one triplet bound state ($T_1$),  and the lower edge  of 
          the continuum ($C_l$) at $k=\pi$,
           for the bound states of two even parity holes  of the $t-J$ ladder with
           $y_1=y_2=0.1$.
Nonzero coefficients $x^n$
up to order $n=11$  are listed.}\label{tab_ser}
\begin{ruledtabular}
\begin{tabular}{rllll} 
\multicolumn{1}{c}{$n$} &\multicolumn{1}{c}{$E_{S_1}/J_{\perp}$ for $S_1$} 
&\multicolumn{1}{c}{$E_{S_2}/J_{\perp}$ for $S_2$}
&\multicolumn{1}{c}{$E_{T_1}/J_{\perp}$ for $T_1$}  &\multicolumn{1}{c}{$E_{C_l}/J_{\perp}$ for $C_l$} \\
\hline
 0 & ~1.8000000000                 & ~1.8000000000                 & ~1.8000000000                 & ~1.8000000000      \\
 1 & ~5.0000000000$\times 10^{-1}$ & ~1.0000000000                 & ~1.0000000000                 & ~1.0000000000      \\
 2 & ~2.2500000000$\times 10^{-1}$ & ~8.5000000000$\times 10^{-1}$ & ~8.0000000000$\times 10^{-1}$ & ~8.5000000000$\times 10^{-1}$  \\
 3 & ~4.2708333333$\times 10^{-1}$ & ~3.2811666667$\times 10^{-1}$ & ~3.8020833333$\times 10^{-1}$ & ~4.5625000000$\times 10^{-1}$  \\
 4 & ~9.8972656250$\times 10^{-1}$ & -5.0189024889$\times 10^{-1}$ & -2.1309606481$\times 10^{-1}$ & -2.0194444444$\times 10^{-1}$  \\
 5 & -1.4465469473$\times 10^{-1}$ & -1.0020098268                 & -6.5367901837$\times 10^{-1}$ & -7.4625737847$\times 10^{-1}$  \\
 6 & -3.6759280211                 & -5.2752447741$\times 10^{-1}$ & -5.1516620601$\times 10^{-1}$ & -5.7308628683$\times 10^{-1}$  \\
 7 & -1.8432595464                 & ~9.3228166551$\times 10^{-1}$ & ~2.9350608457$\times 10^{-1}$ & ~4.7837543431$\times 10^{-1}$  \\
 8 & ~1.2540109894$\times 10^{1}$  & ~2.4273057426                 & ~1.1202826230                 & ~1.5454790111      \\
 9 & ~1.1255237404$\times 10^{1}$  & ~2.2797858202                 & ~8.6634384749$\times 10^{-1}$ & ~1.0984536226      \\
10 & -4.7671925391$\times 10^{1}$  & -5.2477491651$\times 10^{-1}$ & -8.6710572784$\times 10^{-1}$ & -1.4110630505      \\
11 & -5.6591915055$\times 10^{1}$  & -4.6480530617                 & -2.6386765415                 & -3.8611918629      \\
\end{tabular}
\end{ruledtabular}
\end{table*}

\begin{table*}
\squeezetable
\caption{$[n/m]$ $D$ log Pad\'e approximants to the series for bining energy $E_b/J_{\perp}$ at $k=\pi$
 of singlet bound state $S_2$ of 4-holes at $t/J=t_{\perp}/J_{\perp}=0.1$.
 An asterisk denotes a defective approximant.
} \label{tab3}
\begin{ruledtabular}
\begin{tabular}{llllll}
\multicolumn{1}{c}{n} &\multicolumn{1}{c}{$[(n-2)/n]$}&\multicolumn{1}{c}{$[(n-1)/n]$}
&\multicolumn{1}{c}{$[n/n]$} &\multicolumn{1}{c}{$[(n+1)/n]$}&\multicolumn{1}{c}{$[(n+2)/n]$}
 \\
\multicolumn{1}{c}{} &\multicolumn{1}{c}{pole (residue)} &\multicolumn{1}{c}{pole (residue)} 
&\multicolumn{1}{c}{pole (residue)} &\multicolumn{1}{c}{pole (residue)} &\multicolumn{1}{c}{pole (residue)} \\
\colrule
 n= 1 &                     &                       & 0.19966(1.1704)     & 0.19987(1.1741)     & 0.19391(1.0401) \\
 n= 2 &                     & 0.19987 ( 1.1741)     & 0.19967(1.1705)     & 0.19194(0.9851)     & 0.19250(1.0032) \\
 n= 3 & 0.19415(1.0485)$^*$ & 0.19255 ( 1.0055)     & 0.19265(1.0088)     & 0.19272(1.0117)     & 0.19280(1.0154) \\
 n= 4 & 0.19264(1.0086)     & 0.19288 ( 1.0202)$^*$ & 0.19296(1.0259)$^*$ & 0.19171(1.0808)$^*$ &                 \\
 n= 5 & 0.19295(1.0259)$^*$ & 0.19285 ( 1.0185)$^*$ &                     &                     &                 \\
\end{tabular}                                                        
\end{ruledtabular}
\end{table*}

\end{document}